\documentclass{article}
\usepackage{epsfig}
\newcommand{\bfr}{\begin{flushright}}
\newcommand{\efr}{\end{flushright}}
 
\begin{document}
% \eqsec  % uncomment this line to get equations numbered by (sec.num)
\title{On Instability of Squashed Spheres in the Kaluza-Klein Theory
%\thanks{Presented at ...}%
% you can use '\\' to break lines
}
\author{Kiyoshi Shiraishi\\
%\address{
National Laboratory for High Energy Physics (KEK)\\
Tsukuba, Ibaraki 305\\
%and\\
Department of Physics, Tokyo Metropolitan University\\
Setagaya, Tokyo 158
%}
}
\date{Prog. Theor. Phys. Vol. 74 No. 4 (1985) pp. 832--841
}
\maketitle
\begin{abstract}
We study in Kaluza-Klein theories stability of the extra space against
``squashing'', in other words, the homogeneous deformation. Quantum
fluctuations of matter fields at one-loop level are taken into
consideration. We calculate the effective potential in models of the
type, $M^4\times S^3$ and $M^4\times S^7$. It is found that in the case
of scalar matter fields the stability depends on the coupling to the
scalar curvature.
\end{abstract}
%\PACS{}

%%%%%%%%%%%%%%%%%%%%%%%%%%%%%%%%%%%%%%%%%%%%%%%%%%%%%%%%%%%%
\section{Introduction}
%%%%%%%%%%%%%%%%%%%%%%%%%%%%%%%%%%%%%%%%%%%%%%%%%%%%%%%%%%%%
Many problems on the unification of interactions in higher dimensions
have been discussed recently.\cite{1} In Kaluza-Klein theories the
ground state is taken to be a product of four dimensional space-time
and some compact homogeneous space whose isometry corresponds to the
gauge symmetry. The length scale associated with the ``extra''
dimensions must be comparable to the Planck length ($\sim 10^{-33}$cm)
in order for gauge couplings in the theory to be of order of
unity.\cite{2} The energy scale of excited modes on the internal space
is therefore the Planck energy ($\sim 10^{19}$ GeV). One of the
difficulties in this approach is how one can find a static compactified
solution of the Einstein equation which has the desired size of the
extra space. Many authors obtained static solutions in models which
include classical bosonic fields and/or fermion condensations. On the
other hand, Candelas and Weinberg~\cite{3} considered quantum effects
of matter fields and showed that there are solutions in which the
background geometry is $M^4\times S^N$. They showed that the number of
matter fields would determine the magnitude of the gauge coupling
constant and the stability against uniform dilatations of the scale of
$S^N$. A special case of the more general spacetime $M^4\times
S^M\times S^N$ is considered by Kikkawa et al.\cite{4} Their stable
solutions due to quantum effects of matter fields give the ratio of
coupling constants. Their model is a proto-type of the so-called
``standard model'' of interactions (i.e., a model of the product
gauge group). Recently Lim~\cite{5} and Okada~\cite{6} discussed the
symmetry breaking in the Kaluza-Klein theories. They found that
symmetries of the isometry group are broken through quantum effects of
(minimally or conformally coupled) scalar matter fields. This symmetry
breaking corresponds to a deformation of the extra space. lt may be
said that their models correspond to grand unified theories which
include spontaneous symmetry breakings.

In this paper, we investigate the stability of extra space against
homogeneous deformation in two cases. In one case, the extra
space is either $S^3$ or $S^7$ with non-minimally coupled scalar
fields, and in the other case the extra space is $S^3$ with Dirac
fermion fields.

The present paper is organized as follows. In \S 2, 1Fe consider
the metric $M^4 \times (\mbox{squashed~} S^3)$ and relation of the gauge
symmetry breaking and the deformation of spheres. In \S 3, we
calculate the one-loop quantum effective potential for the metric of
$M^4\times S^3$ and discuss the stability against ``squashing''. The
stability for the background geometry of $M^4\times S^7$ is also
discussed in \S 4. The last section is devoted to discussion.

%%%%%%%%%%%%%%%%%%%%%%%%%%%%%%%%%%%%%%%%%%%%%%%%%%%%%%%%%%%%
\section{Homogeneous deformations of $S^3$}
%%%%%%%%%%%%%%%%%%%%%%%%%%%%%%%%%%%%%%%%%%%%%%%%%%%%%%%%%%%%
The symmetry of $S^3$ is well known in particular through the
investigation of the mixmaster universe model. We can express a line
element of three dimensional space as follows:
\begin{equation}
d\ell^2=a^2\sigma_1^2+b^2\sigma_2^2+c^2\sigma_3^2
\label{2.1}
\end{equation}
and
\begin{eqnarray}
\sigma_1&=&-\sin\psi d\theta+\cos\psi\sin\theta d\phi\,,\nonumber \\
\sigma_2&=&\cos\psi d\theta+\sin\psi\sin\theta d\phi\,,\nonumber \\
\sigma_3&=&d\psi+\cos\theta d\phi\,.\nonumber
\end{eqnarray}
Here $a$, $b$ and $c$ are scale factors, and in the case $a=b=c$, this
line element corresponds to that of a maximally symmetric 3-sphere.

When three scale factors take different values, this space has lower
symmetry. We will denote this deformable space as $\hat{S}^3$ or
``squashed'' $S^3$.

If one uses this space as the extra space in the Kaluza-Klein
theory, the gauge
symetry breaking can be discussed. To see this, we consider the
seven dimensional
geometry as (Kaluza-Klein ansatz):
\begin{equation}
ds^2=\eta_{\mu\nu}dx^\mu dx^\nu+a^2 s_1^2+b^2 s_2^2+c^2 s_3^2
\label{2.2}
\end{equation}
and $s_i=\sigma_i+K_i^\alpha A_\mu^\alpha(x)dx^\mu$, ($i=1,2,3;
\alpha=1,2,\dots, 6$) where $\eta_{\mu\nu}={\rm diag}(-1,1,1,1)$,
$x^\mu$ are coordinates of four dimensional flat space, and $K_i^\alpha$
are $i$-th components of six Killing vectors on $S^3$. The isometry
group of $S^3$ is $SO(4)$, which has six generators.

In this case, the four dimensional effective action of gauge fields 
after being reduced from the seven dimensional Einstein-Hilbert
action is given by \cite{7}
\begin{eqnarray}
& &\int d^4x
\left[-\frac{1}{4}\left\{a^2(F^1_{\mu\nu})^2+b^2(F^2_{\mu\nu})^2+
c^2(F^3_{\mu\nu})^2\right.\right.\nonumber
\\
& &\quad\left.\left.+\frac{1}{3}(a^2+b^2+c^2)
\{(F^4_{\mu\nu})^2+
(F^5_{\mu\nu})^2+(F^6_{\mu\nu})^2\}\right\}\right.\nonumber \\ &
&\quad-\left.\frac{1}{2}\left\{\frac{(b^2-c^2)^2}{b^2c^2}(A^1_{\mu})^2
+\frac{(c^2-a^2)^2}{c^2a^2}(A^2_{\mu})^2
+\frac{(a^2-b^2)^2}{a^2b^2}(A^3_{\mu})^2\right\}\right]\,.  
\label{2.3}
\end{eqnarray}

There appear mass terms of gauge bosons in general. From (\ref{2.3})
gauge symmetries are shown to be
\begin{eqnarray}
SU(2)\times SU(2)\sim SO(4)& & \mbox{when~} a=b=c\,,\nonumber \\
SU(2)\times U(1)& &  \mbox{when~} a=b\ne c
\quad\mbox{etc}\,,\nonumber
\\SU(2)& & \mbox{when~} a\ne b\ne c\,.\nonumber
\end{eqnarray}
The gauge symmetry breaking of this type is extensively investigated by
Okada.\cite{6} He found that the quantum effect of conformally coupled
scalar field would break the symmetry,

We shall consider here only the possibility of breaking of maximal
symmetry, that is, the stability of ``round'' sphere, for simplicity.
However, we deal with quantum effects of non-minimally coupled
scalar fields generally, as well as fermion matter fields.

%%%%%%%%%%%%%%%%%%%%%%%%%%%%%%%%%%%%%%%%%%%%%%%%%%%%%%%%%%%%
\section{Stability of $\hat{S}^3$}
%%%%%%%%%%%%%%%%%%%%%%%%%%%%%%%%%%%%%%%%%%%%%%%%%%%%%%%%%%%%
In order to calculate one-loop quantum effects, we must know the
spectrum of the wave operator on $\hat{S}^3$.

First, let us consider the scalar field coupled to gravity
nonminimally as the matter field. The Lagrangian density (for
matter+gravity) is
\begin{equation}
L=-\frac{1}{2}(\partial_M\Phi)^2+\frac{1}{2}\xi
R\Phi^2-\frac{1}{16\pi\bar{G}}(R+2\lambda)\,,   
\label{3.1}
\end{equation}
where $R$ is the scalar curvature.

The scalar boson mass matrix on $\hat{S}^3$ can be written as
\begin{equation}
M^2=\frac{1}{a^2}L_1^2+\frac{1}{b^2}L_2^2+\frac{1}{c^2}L_3^2
-\xi\tilde{R}\,,    
\label{3.2}
\end{equation}
where
\begin{equation}
\tilde{R}=-\frac{1}{2a^2b^2c^2}(2b^2c^2+2c^2a^2+2a^2b^2-a^4-b^4-c^4)\,.
\end{equation}

Operators $L_1, L_2$ and $L_3$ satisfy the same algebraic relation as
the angular momentum.\cite{8} When $a=b$, the mass matrix can be
diagonalized as
\begin{equation}
M^2_{(L,m)}=\frac{1}{a^2}\{L(L+1)-m^2\}+\frac{1}{c^2}m^2-\xi\tilde{R}    
\label{3.3}
\end{equation}
with $L=0,1/2,1,\dots, m=-L,-L+1,\dots,L$.

We use the dimensional regularization to calculate the effective
potential $V_1$ as in Ref.~\cite{3}:
\begin{equation}
V_1=-\frac{1}{2(4\pi)^2}\Gamma\left(-\frac{n}{2}\right){\rm Tr~} 
D\,(M^2)^{n/2}\,,\qquad n\rightarrow 4\,,
\label{3.4}
\end{equation}
where $D$ is the degeneracy of the states. In our case, $D=2L+1$. The
one-loop effective potential for $M^4\times\hat{S}^3$ is then
given by
\begin{equation}
V_1=-\frac{1}{2(4\pi)^2}\frac{1}{(2a)^4}\Gamma\left(-\frac{n}{2}\right)
\sum_{l=1}^\infty\,l\,\sum_m\left[l^2-1+4m^2\left(\frac{a^2}{c^2}-1
\right)-\xi\tilde{R}\cdot(2a)^2\right]^{n/2}\,,
\label{3.5}
\end{equation}
when $a=b$. The summation on $m$ is taken for $m=-(l-1)/2,-(l-3)/2,
\dots,(l-1)/2$.

In the general case ($a\ne b\ne c$) , let us use the following
parametrization:
\begin{eqnarray}
a&=&\bar{a}\exp\left(u+\frac{\sqrt{5}}{2}v+\frac{\sqrt{15}}{2}w\right)
\,,\nonumber
\\
b&=&\bar{a}\exp\left(u+\frac{\sqrt{5}}{2}v-\frac{\sqrt{15}}{2}w\right)
\,,\nonumber\\
c&=&\bar{a}\exp\left(u-{\sqrt{5}}v\right)
\,.
\label{3.6}
\end{eqnarray}
$v$ or $w$ represents presence of a nonvanishing deformation from the
sphere, keeping the volume of extra space constant. Following
Moss,\cite{9} we choose coordinates in which the metric takes the form
\begin{equation}
ds^2=e^{-3u}\eta_{\mu\nu}dx'{}^\mu dx'{}^\nu+dl^2\,.
\label{3.7}
\end{equation}

The effective potential (including the tree level potential) can be
written as
\begin{eqnarray}
U&=&\frac{2}{15}e^{-3u}\left[\lambda-\frac{1}{(2\bar{a})^2}e^{-2u}
\{4e^{-\sqrt{5}v}\cosh\sqrt{15}w\right.\nonumber \\
&
&\left.-e^{-4\sqrt{5}v}-4e^{2\sqrt{5}v}(\sinh\sqrt{15}w)^2\}+8\pi\bar{G}
\frac{J(v, w)}{2\pi^2(2\bar{a})^7}e^{-7u}\right]\,.
\label{3.8}
\end{eqnarray}
The first term includes the cosmological constant $\lambda$ and the
second term comes from the curvature of the extra space. The last term
includes quantum effects, while $J(v, w)$ is defined as
\begin{equation}
V_1=\frac{e^{-4u}}{(2\bar{a})^4}J(v,w)\,.
\label{3.9}
\end{equation}
Then Einstein equations give
\begin{eqnarray}
\frac{1}{2}\{(\partial_\mu u)^2+(\partial_\mu
v)^2+(\partial_\mu w)^2\}+U&=&0\,,
\nonumber \\
-\partial^\mu\partial_\mu u+\frac{\partial U}{\partial u}&=&0\,,
\nonumber \\
-\partial^\mu\partial_\mu v+\frac{\partial U}{\partial v}&=&0\,,
\nonumber \\
-\partial^\mu\partial_\mu w+\frac{\partial U}{\partial w}&=&0\,.
\label{3.10}
\end{eqnarray}
Therefore, in order to obtain the static stable solution without
deformation from the spherical symmetry ($v=w=0$), we can choose
$\lambda$ and $\bar{a}$ such that $U(u=v=w=0)=\partial U
/\partial u(u=v=w=0))=0$ and
$\partial^2 U/\partial u^2(u=v=w=0)>0$ provided that
$J(v=w=0)\equiv J_0>0$. Our choice is
\begin{equation}
\lambda=\frac{8\pi\bar{G}}{2\pi^2(2\bar{a})^7}\frac{5}{2}J_0\,,
\quad
\frac{6}{(2\bar{a})^2}=\frac{8\pi\bar{G}}{2\pi^2(2\bar{a})^7}
\cdot 7J_0\,.
\label{3.11}
\end{equation}

Here, we consider the stability against the defomations. It is easily
found that $\partial U/\partial v=\partial U/\partial w=0$ and
$\partial^2U/\partial v^2=\partial^2U/\partial w^2$ at
$u=v=w=0$ hold in our model (see Appendix A). The stability condition
against the perturbation of $v$ is then 
\begin{equation}
\frac{\partial^2U}{\partial v^2}(u=v=w=0)=
\frac{2}{15}\frac{8\pi\bar{G}}{2\pi^2(2\bar{a})^7}\delta>0\,,
\label{3.12}
\end{equation}
where
\[
\delta\equiv 70J_0+\frac{\partial^2J}{\partial v^2}(v=w=0)\,.
\]

\begin{figure}[ht]
\begin{center}
\includegraphics[width=6cm]{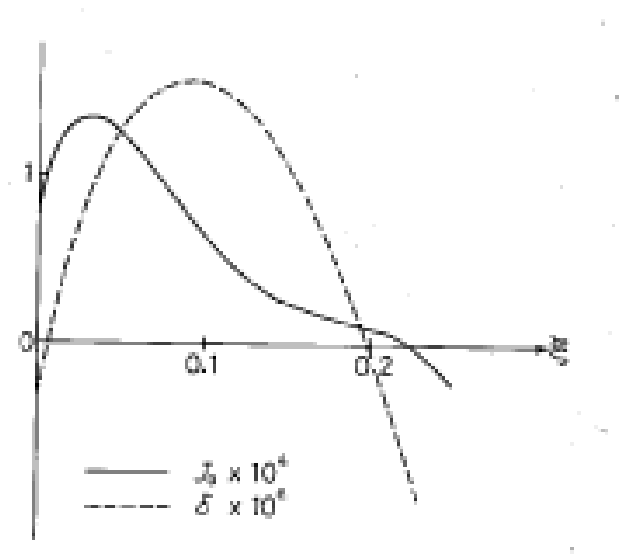}
\caption{$J_0$ and $\delta$ are plotted against $\xi$ in the case
$M^4\times \hat{S}^3.$}
\label{f2}\end{center}
\end{figure}

The numerical calculation of the effective potential is performed by the
method shown in Appendix A. Figure~\ref{f1} shows $J(v)$; $J(v,w=0)$
plotted against $v$. We show $\delta$ and $J_0$ against the
scalar-curvature coupling $\xi$ in Fig.~\ref{f2}. It is found that the
maximally symmetric $\hat{S}^3$ can be stabilized when $0.007\le\xi
\le 0.198$.

%**************************************************************
\begin{figure}[ht]
\begin{center}
\includegraphics[width=6cm]{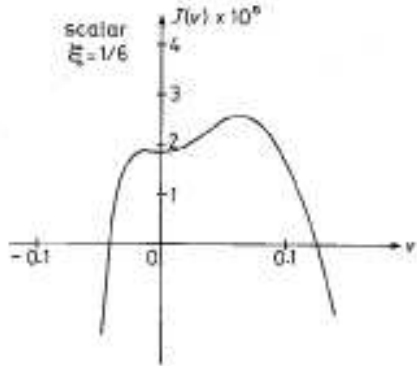}
\caption{The result of the numerical calculations for
$J(v)$ due to a scalar field ($\xi=1/6$) in the case $M^4\times
\hat{S}^3$.}
\label{f1}\end{center}
\end{figure}
%**************************************************************

%**************************************************************
%**************************************************************

We also calculate the quantum effect of Dirac fermion field (see
Appendix A). $J(v)$ is shown in Fig.~\ref{f3}. We find
\[
\delta\approx -0.176<0.
\]

%**************************************************************
\begin{figure}[ht]
\begin{center}
\includegraphics[width=6cm]{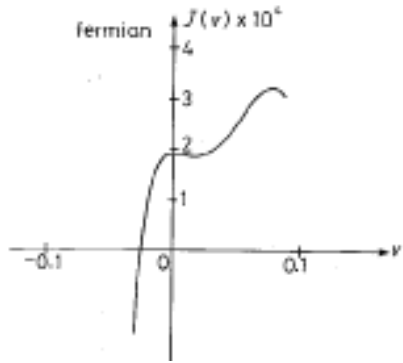}
\caption{The result of the numerical calculations for $J(v)$ due to a
fermion field in the case $M^4\times\hat{S}^3$.}
\label{f3}\end{center}
\end{figure}
%**************************************************************

From (\ref{3.12}), it is impossible to stabilize $S^3$ with the quantum
effect of Dirac fermion fields only.

%%%%%%%%%%%%%%%%%%%%%%%%%%%%%%%%%%%%%%%%%%%%%%%%%%%%%%%%%%%%
\section{The case $M^4\times\hat{S}^7$}
%%%%%%%%%%%%%%%%%%%%%%%%%%%%%%%%%%%%%%%%%%%%%%%%%%%%%%%%%%%%
It is well known that $S^7$ is deformed homogeneously, or
``squashed''.\cite{10} The deformable $S^7$ has the following line
element:
\begin{equation}
dl^2=a^2\left(d\mu^2+\frac{1}{4}\sin^2\mu\sum_{i=1}^3\omega_i^2
\right)+\frac{1}{4}\sum_{i=1}^3b_i^2\left(
\nu_i+\cos\mu\omega_i\right)^2\,,   
\label{4.1}
\end{equation}
where
\[
\nu_i=\sigma_i+\Sigma_i\,,\quad \omega_i=\sigma_i-\Sigma_i
\]
and they satisfy the algebra, such as
\[
d\sigma_1=-\sigma_2\wedge\sigma_3\,,\quad d\Sigma_1=-\Sigma_2\wedge
\Sigma_3\,.
\]
The scalar curvature is
\begin{equation}
-\left[\frac{12}{a^2}-\frac{b_1^2+b_2^2+b_3^2}{a^4}+
\frac{1}{2b_1^2b_2^2b_3^2}(2b_2^2b_3^2+2b_3^2b_1^2+2b_1^2b_2^2
-b_1^4-b_2^4-b_3^4)\right]\,.
\label{4.2}
\end{equation}
For simplicity, we take $b_1=b_2=b_3=b$.

Similarly to the last section, let us parametrize $a$ and $b$ as
\begin{eqnarray}
a&=&\bar{a}\exp\left(u+\frac{3}{2}\sqrt{\frac{3}{2}}v\right)\,,\nonumber
\\ b&=&\bar{a}\exp\left(u-2\sqrt{\frac{3}{2}}v\right)\,.
\end{eqnarray}

The mass spectrum on $\hat{S}^7$ was already given by Nilsson and
Pope.\cite{11} We calculate the quantum effect of nonminimally coupled
scalar fields in the similar way as in the last section (see Appendix
B). The result is shown in shown in Fig.~\ref{f4}. Here $\beta$
indicates the effective four dimensional Newton constant
(see Appendix B), as given by Ref.~\cite{3}. Then we consider that the
region of $\xi$ in which $\beta$ is negative has no physical meaning.

%**************************************************************
\begin{figure}[ht]
\begin{center}
\includegraphics[width=6cm]{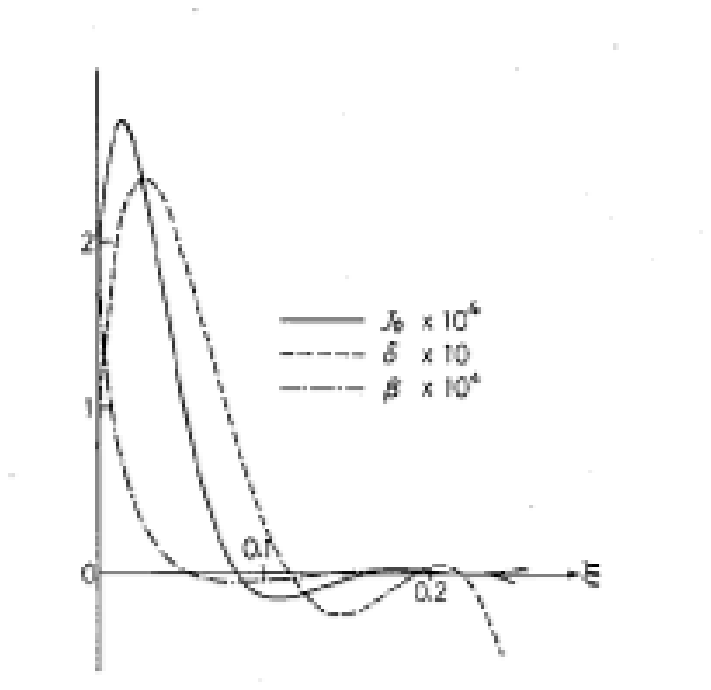}
\caption{$J_0$, $\delta$ and $\beta$ are plotted against $\xi$ in the
case $M^4\times\hat{S}^7$.}
\label{f4}\end{center}
\end{figure}
%**************************************************************

We find that in order to stabilize $S^7$, $\xi$ must fall in a region
given by $0<\xi\le 0.054$ or $0.193\le\xi\le 0.218$.

%%%%%%%%%%%%%%%%%%%%%%%%%%%%%%%%%%%%%%%%%%%%%%%%%%%%%%%%%%%%
\section{Discussion}
%%%%%%%%%%%%%%%%%%%%%%%%%%%%%%%%%%%%%%%%%%%%%%%%%%%%%%%%%%%%
In the present paper, we have investigated the stability of deformable
spheres. It is shown that the stability depends on the coupling $\xi$ to
the scalar curvature in the case that the quantum effect of scalar
fields is taken into account.

We have studied very limited types of deformations. The structures of
the deformable spheres have mathematically interesting features. It
is also interesting that $M^4\times\mbox{(extra space)}$ is regarded as
the result of some sorts of deformations, since we study a spontaneous
compactification of a dimensional reduction as an effect of dynamical
time evolution of scale factors in the cosmological context. In
addition, deformations of extra spaces may require some modification
in the scenario of the ``Kaluza-Klein Inflation \cite{12}''.

%%%%%%%%%%%%%%%%%%%%%%%%%%%%%%%%%%%%%%%%%%%%%%%%%%%%%%%%%%%%
\section*{Acknowledgements}
%%%%%%%%%%%%%%%%%%%%%%%%%%%%%%%%%%%%%%%%%%%%%%%%%%%%%%%%%%%%
The author would like to thank M.~Yoshimura for critical reading of
the manuscript.

%%%%%%%%%%%%%%%%%%%%%%%%%%%%%%%%%%%%%%%%%%%%%%%%%%%%%%%%%%%%
\section*{Appendix A}
%%%%%%%%%%%%%%%%%%%%%%%%%%%%%%%%%%%%%%%%%%%%%%%%%%%%%%%%%%%%
We give the details of calculations of the effective potential for the
model, the geometry of which is $M^4\times\hat{S}^3$.

\smallskip
\noindent
{\it Scalar fields}

We rewrite $J$ as
\begin{equation}
J(v, w)=-\frac{e^{-2\sqrt{5}v}}{2(4\pi)^2}\Gamma\left(-\frac{n}{2}
\right)\sum_{l=1}^\infty\,l {\rm tr~}[M_0^2+m_1^2]^{n/2}\,,\quad
n\rightarrow 4\,, 
\label{A.1}
\end{equation}
where
\begin{eqnarray}
M_0^2&=&\cosh\sqrt{15}w\,(l^2-1)+2\xi\{4\cosh\sqrt{15}w-
e^{-3\sqrt{5}v}-4e^{-3\sqrt{5}v}(\sinh\sqrt{15}w)^2\}\,,
\nonumber \\
m_1^2&=&4(e^{3\sqrt{5}v}-\cosh\sqrt{15}w)L_3^2-2\sinh\sqrt{15}w
(L_+^2+L_-^2)\nonumber
\end{eqnarray}
and
\[
L_+=L_1+iL_2\,,\quad L_-=L_1-iL_2\,.
\]
Therefore,
\[
J_0=J(0,0)=-\frac{1}{2(4\pi)^2}\Gamma\left(-\frac{n}{2}\right)
\sum_{l=1}^\infty l^2 (l^2-1+6\xi)^{n/2}\,,\quad n\rightarrow 4\,.
\]
This agrees with $C_3^{(0)}$ ($C_3^{conformal}$) in Ref.~\cite{3} when
$\xi=0$ ($\xi=5/24$).

The expressions for $\partial J/\partial v, \partial J/\partial w,
\partial^2J/\partial v^2, \partial^2J/\partial w^2$ and
$\partial^2J/\partial v\partial w$ is derived from the perturbative
method. They are found to be
\begin{eqnarray}
\frac{\partial J}{\partial v}(v=w=0)&=&\frac{\partial J}{\partial
w}(v=w=0)=0\,,\nonumber \\ \frac{\partial^2J}{\partial
v^2}(v=w=0)&=&\frac{\partial^2J}{\partial
v^2}(v=w=0)\nonumber \\
&=&20J_0-\frac{4}{(4\pi)^2}\Gamma\left(-\frac{n}{2}\right)
\sum_l l^2(l^2-1)(l^2-4)(l^2-1+6\xi)^{n/2-2}\nonumber \\
& &+\frac{180}{(4\pi)^2}\Gamma\left(-\frac{n}{2}\right)\sum_l l^2
(l^2-1+6\xi)^{n/2-1}\,,\nonumber \\
\frac{\partial^2J}{\partial v\partial w}(v=w=0)&=&0\,.           
\label{A.2}
\end{eqnarray}

In order to evaluate these quantities, we use the following identity:
\begin{equation}
\Gamma\left(-\frac{n}{2}\right)(M^2)^{n/2}=\int_0^\infty dt\,
t^{-n/2-1}\exp(-t M^2)\,.
\label{A.3}
\end{equation}
For example, it leads to
\begin{equation}
J_0=-\frac{1}{32\pi^2}\sum_{l=1}^\infty l^2\int dt\,t^{-n/2-1}
\exp\{-t(l^2-1+6\xi)\}\,.
\label{A.4}
\end{equation}
Further, using the reaection formula \cite{13}
\begin{equation}
\sum_{l=1}^\infty l^2 \exp(-l^2t)=\frac{\sqrt{\pi}}{t^{3/2}}
\left[\sum_{l=1}^\infty\left(\frac{1}{2}-\frac{\pi^2 l^2}{t}
\right)\exp\left(-\frac{\pi^2l^2}{t}\right)+\frac{1}{4}\right]\,,
\label{A.5}
\end{equation}
it can be shown as
\begin{eqnarray}
J_0&=&-\frac{\sqrt{\pi}}{32\pi^2}\int_0^\infty dt\,t^{-n/2-5/2}\exp\{
-t(6\xi-1)\}\left\{\sum_{l=1}^\infty\left(\frac{1}{2}-\frac{\pi^2l^2}{t}
\right)e^{-\frac{\pi^2l^2}{t}}+\frac{1}{4}\right\}
\nonumber \\
&=&-\frac{\sqrt{\pi}}{32\pi^2}\left[\sum_{l}
\left\{\frac{(6\xi-1)^{1/2}}{\pi
l}
\right\}^{7/2}K_{7/2}(2(6\xi-1)^{1/2}\pi l)\right.
\nonumber \\
& &-\sum_l 2\pi^2 l^2\left\{\frac{(6\xi-1)^{1/2}}{\pi
l}
\right\}^{9/2}K_{9/2}(2(6\xi-1)^{1/2}\pi
l)\nonumber \\
&
&\left.+\frac{1}{4}\Gamma\left(-\frac{7}{2}\right)(6\xi-1)^{7/2}\right]
\,.
\label{A.6}
\end{eqnarray}
In the last line of (\ref{A.6}) , we have used the identity including
the modified Bessel function $K_\nu(z)$:
\begin{equation}
K_\nu(z)
=\frac{1}{2}\left(\frac{z}{2}\right)^\nu
\int_0^\infty\exp\left(-t-\frac{z^2}{4t}\right)t^{-\nu-1} dt\,.    
\label{A.7}
\end{equation}
Finally, we obtain
\begin{equation}
J_0=-\frac{1}{840\pi}(6\xi-1)^{7/2}+\frac{1}{32\pi}(6\xi-1)^2
\sum_{l=1}^\infty\frac{e^{-z}}{\pi^3
l^3}\left(1+\frac{9}{z}+\frac{39}{z^2}+\frac{90}{z^3}+
\frac{90}{z^4}\right)\,,
\label{A.8}
\end{equation}
where $z=2(6\xi-1)^{1/2}\pi l$.

The value of $\partial^2J/\partial v^2 (v=w=0)$ is obtained in a
similar way. The result is
\begin{eqnarray}
\partial^2J/\partial v^2 (v=w=0)&=&\frac{1}{12\pi}(6\xi-1)^{3/2}
\{(6\xi-1)^2-2\}\nonumber \\
&&+\frac{7}{8\pi}(6\xi-1)^2
\sum_{l=1}^\infty\frac{e^{-z}}{\pi^3
l^3}\left(1+\frac{9}{z}+\frac{39}{z^2}+\frac{90}{z^3}+
\frac{90}{z^4}\right)\nonumber \\
&&+\frac{3}{8\pi}(19\xi+1)(6\xi-1)^{3/2}
\sum_{l=1}^\infty\frac{e^{-z}}{\pi^2
l^2}\left(1+\frac{5}{z}+\frac{12}{z^2}+\frac{12}{z^3}\right)\nonumber \\
&&+\frac{9}{4\pi}\xi(2\xi+1)(6\xi-1)
\sum_{l=1}^\infty\frac{e^{-z}}{\pi
l}\left(1+\frac{2}{z}+\frac{2}{z^2}\right)\,.
\label{A.9}
\end{eqnarray}

To draw Fig.~\ref{f1}, we use the method discussed in Appendix D of
Ref.~\cite{3}. At the first step, we expand as
\begin{equation}
\Gamma\left(-\frac{n}{2}\right)(A+B)^{n/2}=\sum_{r=0}^\infty
\frac{\Gamma(r-n/2)}{r!}
A^{n/2-r}B^r(-1)^r\,.       
\end{equation}
Next, we calculate the finite sum on $m$, and use the identity
\begin{equation}
\Gamma\left(\frac{1}{2}z\right)\zeta(z)=
\pi^{z-1/2}\Gamma\left(\frac{1-z}{2}\right)\zeta(1-z)\,.
\label{A.11}
\end{equation}
The behavior in the vicinity of $v=0$ agrees with the result obtained
by (\ref{A.9}). 

\smallskip
\noindent
{\it Fermions}

The mass spectrum is shown by Dowker \cite{14} as
\begin{equation}
M_\pm=\frac{1}{2a}\left\{\frac{1}{2}\frac{c}{a}\pm\sqrt{(2L+1)^2+
4m^2\left(\frac{a^2}{b^2}-1\right)}\right\}\quad\mbox{when~} a=b
\label{A.12}
\end{equation}
with $-(L\pm 1/2)\le m\le L\pm 1/2$ and $L\ge 0$ for $M_+$ and $L>1/2$
for $M_-$. The degeneracy is $2L+1$.

Consequently, we find
\begin{eqnarray}
J(v))=J(v, 0)
&=&\frac{4}{2(4\pi)^2}e^{4\sqrt{5}v}\Gamma\left(-\frac{n}{2}\right)
\left[\sum_{l=1}^\infty l
\sum_{q=0}^l\left\{\sqrt{l^2+4\gamma
q(l-q)}+\frac{1}{2}e^{-3\sqrt{5}v}\right\}^n\right.\nonumber
\\ & &+\sum_{l=1}^\infty l
\left.\sum_{q=1}^{l-1}\left\{\sqrt{l^2+4\gamma
q(l-q)}-\frac{1}{2}e^{-3\sqrt{5}v}\right\}^n\right]\nonumber \\
&=&\frac{1}{4\pi^2}e^{4\sqrt{5}v}\left[\sum_{r=0}^\infty\frac{\Gamma(
2r-n)}{\Gamma(-r)(2r)!}\Gamma\left(-\frac{n}{2}\right)
\right.\nonumber \\
& &\times\sum_{l=1}^\infty l
\sum_{q=0}^l[l^2+4\gamma q(l-q)]^{n/2-r}\left(\frac{1}{2}
e^{-3\sqrt{5}v}\right)^{2r}\nonumber \\
&
&\left.-\sum_{r=0}^\infty\frac{\Gamma(2r+1-n)}{\Gamma(-n)(2r+1)!}\Gamma
\left(-\frac{n}{2}\right)\zeta(2r-n)\left(\frac{1}{2}
e^{-3\sqrt{5}v}\right)^{2r+1}\right]\nonumber
\end{eqnarray}
with
\[
\gamma\equiv e^{-3\sqrt{5}v}-1\,.
\]
Further expansions enable us to evaluate $J(v)$.

%%%%%%%%%%%%%%%%%%%%%%%%%%%%%%%%%%%%%%%%%%%%%%%%%%%%%%%%%%%%
\section*{Appendix B}
%%%%%%%%%%%%%%%%%%%%%%%%%%%%%%%%%%%%%%%%%%%%%%%%%%%%%%%%%%%%
Here we deal with the calculation for the model $M^4\times\hat{S}^7$.

Using the mass spectrum on $\hat{S}^7$ given by Nilsson and
Pope,\cite{11} we obtain
\begin{eqnarray}
& &J(v)=e^{4u}(2\bar{a})^4V_1\nonumber \\
& &=-\frac{1}{2(4\pi)^2}e^{-6\sqrt{3/2}v}\Gamma\left(-\frac{n}{2}
\right)\frac{1}{48}\sum_{l=0}^\infty(l+3)\sum_{q}q^2\{(l+3)^2-q^2\}
\nonumber \\
&
&\times[(l+3)^2-9+(e^{7\sqrt{3/2}v}-1)(q^2-1)+6\xi(e^{7\sqrt{3/2}v}
+8-2e^{-7\sqrt{3/2}v})]^{n/2}   
\label{B.1}
\end{eqnarray}
where $q=-(l+1),-(l+1)+2,\dots,l+1$.

Of course, $J_0$ agrees with $C_7^{(0)}$ in Ref.~\cite{3} when $\xi=0$,
and at
$v=0$ are perturbatively evaluated similarly to Appendix A.

The effective potential in this case is
\begin{equation}
U=\frac{2}{63}e^{-7u}\left[\lambda-\frac{3}{(2\bar{a})^2}e^{-2u}\{
8 e^{-3\sqrt{3/2}v}-2e^{-10\sqrt{3/2}v}+e^{4\sqrt{3/2}v}\}+8\pi\bar{G}
\frac{J(v) e^{-11u}}{(\pi^4/3)(2\bar{a})^{11}}\right]\,.
\label{B.2}
\end{equation}
The stability conditions against $u$ are
\begin{equation}
\lambda=\frac{8\pi\bar{G}}{(\pi^4/3)(2\bar{a})^{11}}\frac{9}{2}J_0\,,
\quad\frac{42}{(2\bar{a})^2}=\frac{8\pi\bar{G}}{(\pi^4/3)(2\bar{a})^{11}}
\cdot 11 J_0\,.
\label{B.3}
\end{equation}
Then,
\[
\frac{\partial^2U}{\partial v^2}(u=v=0)=\frac{2}{63}
\frac{8\pi\bar{G}}{(\pi^2/3)(2\bar{a})^{11}}\delta
\]
with
\begin{equation}
\delta\equiv 132J_0+\frac{\partial^2J}{\partial v^2}(v=0)\,.     
\label{B.4}
\end{equation}

$\beta$ is defined by \cite{3,15}
\[
\beta=\frac{11}{42}J_0+2E\quad\mbox{for~} S^7
\]
and
\begin{equation}
E
=\frac{1}{192\pi^2}(1-6\xi)\Gamma\left(-\frac{n}{2}\right)\sum_{l=1}^\infty
\frac{l^2(l^2-1)(l^2-4)}{360}(l^2-9+42\xi)^{n/2-1}\,.
\label{B.5}
\end{equation}
The inverse of effective Newton constant is proportional to $\beta$.

For the case $M^4\times S^3$, $\beta>0$ in the region of $\xi$ that
$J_0>0$.

%%%%%%%%%%%%%%%%%%%%%%%%%%%%%%%%%%%%%%%%%%%%%%%%%%%%%%%%%%%%

%%%%%%%%%%%%%%%%%%%%%%%%%%%%%%%%%%%%%%%%%%%%%

\begin{thebibliography}{99}
%%%%%%%%%%%%%%%%%%%%%%%%%%%%%%%%%%%%%%%%%%%%%%%%%%%%%%%%%%%%
\bibitem{1} H. C. Lee, {\it An Introduction to Kaluza-Klein Theories}
(World Scientific, Singapore, 1984), and references therein.
\bibitem{2} S. Weinberg, Phys. Lett. {\bf B125} (1983) 265.
\bibitem{3} P. Candelas and S. Weinberg, Nucl. Phys. {\bf B237} (1984)
397.
\bibitem{4} K. Kikkawa, T. Kubota, S. Sawada and M. Yamasaki,
Phys. Lett. {\bf B144} (1984) 365.
\bibitem{5} C. S. Lim, Phys. Rev. {\bf D31} (1985) 2507.
\bibitem{6} J. Okada, Class. Quant. Grav. {\bf 3} (1986) 221.
\bibitem{7} R. Coquereaux, Acta Phys. Pol. {\bf B15} (1984) 821.
\bibitem{8} B. L. Hu, S. A. Fulling and L. Parker, Phys. Rev. {\bf D8}
(1973) 2377.
\bibitem{9} I. G. Moss, Phys. Lett. {\bf B140} (1984) 29.
\bibitem{10} M. A. Awada, M. J. Duff and C. N. Pope, Phys. Rev. Lett.
{\bf 50} (1983) 294.
\bibitem{11} B. Nilsson and C. N. Pope, Phys. Lett. {\bf B133} (1983)
67.
\bibitem{12} D. Bailin, A. Love and J. Stein-Schabes, Nucl. Phys.
{\bf B253} (1985) 387.
\bibitem{13} M. Yoshimura, Phys. Rev. {\bf D30} (1984) 344.
\bibitem{14} J. S. Dowker, in {\it Quantum Theory of Gravity} (Adam
Hilger, Bristol, 1984).
\bibitem{15} M. A. Awada and D. J. Toms, Nucl. Phys. {\bf B245} (1984)
161.
\end{thebibliography}
\end{document}